# Anapole Correlations in $Sr_2IrO_4$ Defy the $j_{eff}$ = 1/2 Model

D D Khalyavin[1] and S W Lovesey[1,2]

[1]ISIS Facility, STFC, Didcot, Oxfordshire OX11 0QX, UK

[2]Diamond Light Source Ltd, Didcot, Oxfordshire OX11 0DE, UK

**Abstract** Zel'dovich (spin) anapole correlations in $Sr_2IrO_4$ unveiled by magnetic neutron diffraction contravene the spin-orbit coupled ground state used by the $j_{eff}$ = 1/2 (pseudo-spin) model. Specifically, spin and space know inextricable knots which bind each to the other in the iridate. The diffraction property studied in the Letter is enforced by strict requirements from quantum mechanics and magnetic symmetry. It has not been exploited in the past, whereas neutron diffraction by anapole moments is established. Entanglement of the electronic degrees of freedom is captured by binary correlations of the anapole and position operators, and hallmarked in the diffraction amplitude by axial atomic multipoles with an even rank.

Dressed, quasiparticle and equivalent operators in quantum mechanics have understandable widespread use. For they all facilitate constructs that capture many-body correlation effects and place them in tractable form. A field theory based on dressed particle operators, as opposed to the usual "bare" particle operators, no longer needs a renormalization procedure and avoids use of nonphysical quantities. In this genre, dressed states are epitomized in solutions of the Jaynes-Cummings model, where they are created by the interaction of the atom and the cavity field and serve as a paradigm of entangled (correlated) quantum systems [1, 2, 3]. In solid state physics, quasiparticle operators appear in many formulations of electronic properties of semi-conductors and metals [4]. They are also prominent in theories of conventional superconductivity through Bogolyubov transformations and formulations of the Barden-Cooper-Schrieffer mechanism. Equivalent operators for electronic degrees of freedom are the same basic concept in another guise. Operators of this type date back to the 1950s when they were introduced for electrons participating in resonance phenomena, e.g., EPR, NMR, and, later, the Mössbauer effect. A publication by Stevens in 1952, on magnetic properties of rare earth ions, proved extremely influential at the time, since when his equivalent, or effective, operators have become a standard tool in theories of magnetic phenomena [5, 6].

Such is the case for the compound of immediate interest, perovskite-type $Sr_2IrO_4$, where striking and unexpected properties emerge from complementary interactions at an atomic level of detail. Indeed, an analogy is made between properties of $Sr_2IrO_4$ and the strange physics of ceramic superconductors (underdoped cuprates or high-$T_c$ materials). To begin with, $Sr_2IrO_4$ is an electrical insulator whereas a vacancy in the electronic valence state suggests the contrary (atomic and magnetic properties of iridium ions in $Sr_2IrO_4$ are gathered in Supplementary Material [28]). The conundrum can be resolved by placing three interactions effecting iridium ions in a solid on a near equal footing: a crystalline electric field, generated by ligands ions, a strong spin-orbit coupling, and strong electron correlations. The resulting electronic configuration can be studied with advantage using a half-integer effective (pseudo-spin) operator created from spin and orbital angular momentum [7, 8]. This correlation of spin and orbital angular momentum promotes a dependence of electronic properties on structural changes. The spin-orbit coupling is a relativistic effect that provides an interaction between the orbital angular momentum (**L**) and electron spin (**S**) in ions, which is expressed by use of a total angular momentum variable **J** = **L** + **S**. The coupling $\propto$ **S** • **L** is safely considered a small perturbation for most discussions of electrons in a solid. However, in heavy elements it need not be weak and indeed has recognizable effects (the coupling increases in magnitude as $Z^4$ to a good approximation, where Z is the atomic number). Influence of electron correlations is enhanced, with a gap in the density of states and an insulating state for $Sr_2IrO_4$ [8]. An *ab initio* study of the compound does not favour a simple Slater insulator, but, instead, one created from substantial cooperation of Mott-type correlation effects [9]. A large rotation of $IrO_6$ octahedra about the c-axis is a distinguishing feature in the compound′s structure, otherwise akin to layered $K_2NiF_4$.

Strictures within the pseudo-spin model of an iridate translate to a major simplification of the amplitude for magnetic neutron diffraction. Fortunately, steps to make the model more realistic trigger radical changes to the amplitude that can be tested. A tetragonal distortion contravenes the model and allows J = 3/2 and J = 5/2 in the Ir ground state (low-spin $Ir^{4+}$ ($5d^5$)-configuration), for example. This modification alone has been shown to produce clear-cut changes to the diffraction amplitude [19]. An impartial approach is to test the diffraction amplitude informed by the relevant magnetic space-group against measured Bragg diffraction patterns [10], and we conclude that the patterns and pseudo-spin model predictions do not match. Specifically, the exact angular anisotropy is delineated and polar (parity-odd) contributions are strictly forbidden in the calculated amplitude by dint of selection rules generated from the space-group. Our method of working is far removed from that of Jeong *et al*. [10], who construct a conventional magnetization-density map from their diffraction patterns. They show that the map does not possess angular anisotropy predicted by the pseudo-spin model, and dispute between the model and diffraction patterns is common ground in the two methods of working. However, Jeong *et al*. [10] base their argument on a simple average of two diffraction patterns gathered with different orientations of the applied magnetic field. Our magnetic symmetry argument implies the averaging is not justified, and an informed analysis of their data that we report confirms it. Moreover, we extract from measured diffraction amplitudes precisely defined atomic entities (multipoles with discrete symmetries) that can be studied in simulations of the electronic structure. In particular, we extract correlations of the electronic Zel′dovich anapole that do not exist in the pseudo-spin model.

It is an advantage to write the amplitude for magnetic neutron diffraction as a sum of electronic multipoles. For one thing, atomic multipoles of the type required to complete the theoretical exercise occur in the interpretation of results obtained with other experimental probes in routine use. We denote an axial (parity-even) multipole of integer rank K by $\langle T^K_Q \rangle$ where projections Q obey $- K \leq Q \leq K$, and angular brackets $\langle ... \rangle$ denote the time-average, or expectation, value of the enclosed spherical operator. The dipole $\langle \mathbf{T}^1 \rangle$ is a linear combination of $\langle \mathbf{S} \rangle$ and $\langle \mathbf{L} \rangle$, to a good approximation. In the forward direction of scattering $\langle \mathbf{T}^1 \rangle = (1/3) \langle 2\mathbf{S} + \mathbf{L} \rangle$. This result, first given by Schwinger [16], makes neutron Bragg diffraction the method of choice for the determination of magnetic structures.

Multipoles of particular interest in our study encapsulate spin and orbital (spatial) degrees of freedom, and they are hallmarked by the fact that their rank is even [17]. Specifically, a quadrupole (K = 2) is the expectation value of ($R_0 \Omega_0$), where $\mathbf{R}$ and $\mathbf{\Omega} = (\mathbf{S} \times \mathbf{R})$ are dipole operators for position and the spin anapole, respectively. The spin anapole was studied by Zel'dovich in the course of investigating parity-violating interactions in electromagnetic theory [26]. Parity-violation in atomic and molecular systems with the observation of electronic anapoles can be traced back to 1974 [29-32], and anapole moments are known to diffract neutrons [27]. Evidently, $\langle T^2_0 \rangle \propto \langle (R_0 \Omega_0) \rangle$ is time-odd (magnetic) and parity-even. A quantum mechanical selection rule forbids even rank multipoles in a J-manifold; specifically, they are forbidden in the pseudo-spin model with J = 5/2 [8, 17, 28].

Neutron diffraction experiments of interest utilized a sample environment with a temperature = 4 K and an applied magnetic field, **H**, with strength up to 5 T (a 5 T magnetic field corresponds to an energy ~ 0.30 meV while the iridium spin-orbit parameter ~ 380 meV) [10]. The resultant field-induced magnetization is described by orthorhombic space groups. Two field directions were employed in the experiments: (I) Ib'c'a (73.551) with Ir ion in sites 8c for **H** // [0, 1, 0], and (II) Fd'd'd (70.530) using sites 16f for **H** // [−1, 1, 0] [18]. Iridium site symmetry is acentric in both magnetic space groups. The weakly ferromagnetic state induced by the field keeps the large spin canting inherited from a zero-field scenario, resulting in the big net moment ~ 0.08$\mu_B$/Ir [10]. This implies swapping the antiferromagnetic dipole component as shown in Fig. 1 (to preserve the antisymmetric exchange) and corresponding change of the magnetic ordering wavevector from (1, 1, 1) to (0, 0, 0). Motifs of allowed quadrupole moments are depicted in Fig. 1. Local Ir coordinates ($\xi$, $\eta$, $\zeta$) are $\xi \propto [0, 0, c]$, $\eta \propto [a, 0, 0]$, $\zeta \propto [0, a, 0]$ for (I), and $\xi \propto [a, a, 0]$, $\eta \propto [0, 0, c]$, $\zeta \propto [a, -a, 0]$ for (II). Note that the $\zeta$-axis coincides with the magnetization direction. A unit vector for the direction of the Bragg wavevector $\mathbf{\kappa} = (\kappa_\xi, \kappa_\eta, \kappa_\zeta)$. Integer Miller indices ($H_o$, $K_o$, $L_o$) for the tetragonal parent structure satisfy ($H_o + K_o + L_o$) even. Henceforth, $L_o = 4n$ with $n$ an integer, and Dirac (parity-odd) multipoles are forbidden with this restriction (a proof is provided in Supplementary Material [28]).

The magnetic amplitude $F_M(\mathbf{\kappa})$ measured with a neutron spin-flip technique is the component of the magnetic amplitude in the direction of the field-induced magnetization [10]. We include in $F_M(\mathbf{\kappa})$ symmetry-allowed dipoles (K = 1), quadrupoles (K = 2) and octupoles (K = 3). According to the magnetic space groups mentioned above, multipoles possess projections Q = 0 ($\zeta$-axis) and ± 2 [28]. The generic result for an abbreviated amplitude informed by magnetic symmetry is purely real [17],

$$F_M(\mathbf{\kappa}) \approx \{\langle (2\mathbf{S} + \mathbf{L})_\zeta \rangle \langle j_0(\kappa) \rangle + \langle L_\zeta \rangle \langle j_2(\kappa) \rangle + (5\kappa_\zeta^2 - 1) \langle T^3_0 \rangle\}$$

$$+ [(\kappa_\xi^2 - \kappa_\eta^2) / (1 - \kappa_\zeta^2)] \{\langle T^2_{+2}\rangle'' + (1 - 3\kappa_\zeta^2) \langle T^3_{+2}\rangle'\}. \quad (1)$$

Diffraction patterns are most often analysed with the simple approximation $F_M \approx [\langle(2\mathbf{S} + \mathbf{L})_\zeta\rangle \langle j_0(\kappa)\rangle]$, where $\kappa = (4\pi) \sin(\theta)/\lambda$ is the magnitude of the Bragg wavevector and $\langle j_0(\kappa)\rangle$ a standard radial integral [10, 17]. The property $\langle j_0(0)\rangle = 1$ leaves the simple $F_M(0)$ equal to the magnetic moment $\langle(2\mathbf{S} + \mathbf{L})_\zeta\rangle$ [16]. Termination of the amplitude (1) at the level of octupoles is usually justified on the grounds that multipoles with ranks $K \geq 4$ are very small in the range of wavevectors of interest, and we find this to be an entirely reasonable approximation to the data in hand. The so-called dipole approximation leaves orbital angular momentum $\langle \mathbf{L}_\zeta\rangle$ as the coefficient of the radial integral $\langle j_2(\kappa)\rangle$ [17]. The quadrupole $\langle T^2_{+2}\rangle''$ is likewise proportional to $\langle j_2(\kappa)\rangle$, while octupoles are a linear combination of $\langle j_2(\kappa)\rangle$ and $\langle j_4(\kappa)\rangle$. We use ′ and ″ to denote real and imaginary parts of multipoles in (1), while a multipole with projection $Q = 0$ is purely real. With $\langle j_n(0)\rangle = 0$ for $n \geq 2$ the amplitude (1) obeys $F_M(0) = \langle(2\mathbf{S} + \mathbf{L})_\zeta\rangle$, and the reported value of the induced moment = 0.08 [10].

The objective is to test the magnetic amplitude (1) against experimental data for field-induced amplitudes in $Sr_2IrO_4$ at a temperature = 4 K [10]. To begin with, the simple approximation $F_M \approx [\langle(2\mathbf{S} + \mathbf{L})_\zeta\rangle \langle j_0(\kappa)\rangle]$ displayed in Fig. 2 returns goodness-of-fits $R_F = 37.66\%$ and $R_F = 44.10\%$ for field directions labelled (I) and (II), respectively. The number of unknowns in (1) is reduced by using $\langle \mathbf{L}_\zeta\rangle \approx 0$, which is expected to be to a good approximation for the orbital component of the induced moment. Moreover, inclusion of $\langle \mathbf{L}_\zeta\rangle$ does not add angular anisotropy to $F_M(\kappa)$ that was found to be extremely large in the high-$\kappa$ reflections, e.g., (4, 2, 0) and (2, 0, 20) with $\sin(\theta)/\lambda \approx 0.41$ Å$^{-1}$ and 0.43 Å$^{-1}$, respectively [10]. Moving ahead, we use the exact representation $\langle T^2_{+2}\rangle'' = [q \langle j_2(\kappa)\rangle]$, and infer a value of the quadrupole parameter q from data. By way of orientation to a significant fit to data we experimented with a parameterization $\langle T^3_0\rangle = [p\, t(\kappa)]$ and $\langle T^3_{+2}\rangle' = [r\, t(\kappa)]$ that is correct within a J-manifold. Tolerable agreement was found with $(q/p) \approx -0.3$ & $-0.5$ for cases (I) and (II), respectively. The common dependence on the Bragg wavevector, $t(\kappa)$, was very different for the two cases, however. To investigate the indication of a difference between field directions more fully, and consolidate results for q, we used exact representations $\langle T^3_Q\rangle' = \{\alpha_f [\langle j_2(\kappa)\rangle + \beta_f \langle j_4(\kappa)\rangle]\}$ with f = 1 & 2 for $Q = 0$ and $Q = +2$, respectively. As we already mentioned, $\beta_1 = \beta_2$ for a J-manifold, while $\beta_1 = (2/9)$ is correct for $J = 5/2$ [17]. The abbreviated amplitude (1) now contains five parameters to be inferred from data. Radial integrals in the fits to data are appropriate for isolated $Ir^{4+}$ (Kobayashi *et al.* [20]) with no attempt on our part to simulate departures due to solid-state effects.

The good fits of the amplitude (1) to 26 measurements displayed in Fig. 2 vindicates its intrinsic merit; $R_F = 12.90\%$ (17.37%) and $R_F = 12.98\%$ (18.71%) for field directions labelled (I) and (II), respectively, and values achieved with $q = 0$ are in brackets. It is beyond reasonable doubt that the quadrupole $\langle T^2_{+2}\rangle'' = [q \langle j_2(\kappa)\rangle]$ is significant for both field directions. A useful measure of its physical importance is the relative roles of $\langle T^2_{+2}\rangle''$ and the diagonal octupole $\langle T^3_0\rangle$ in fits to data, and inferred ratios $q/\alpha_1 \approx -0.10$ & $-0.21$ for (I) and (II) are similar to those retrieved by experimenting with a common dependence $t(\kappa)$. Such is our finding for case (I) with $\beta_1 = \beta_2 \approx -2.294$. Inferred values of $\beta_f$ bracket $\approx -1.9$ and $\approx -2.3$ and emphatically rule against use of the $J = 5/2$ manifold. The quantities $[\langle j_2(\kappa)\rangle + \beta\langle j_4(\kappa)\rangle]$ in Fig. 3 are radial dependences of octupoles in (1) for $\beta = (2/9)$ ($J = 5/2$) and $\beta_f$ inferred from data for cases (I) and (II). It is worth noting that an improvement to (1) admits two hexadecapoles, $\langle T^4_{+2}\rangle''$ and $\langle T^4_{+4}\rangle''$ proportional to $\langle j_4(\kappa)\rangle$ [17, 19], that will increase the number of parameters to seven ($\langle T^4_0\rangle$ is forbidden by site symmetry).

In summary, we have exposed binary correlations of anapole and position operators in $Sr_2IrO_4$ that do not exist in the $j_{eff} = 1/2$ (pseudo-spin) model of the spin-orbit coupled ground state. (Anapoles are known to be essential entities in the science of a raft of materials, including magneto-electrics and high-$T_c$ superconducting materials [21, 22, 23].) Empirical evidence for the correlations is derived from neutron Bragg diffraction patterns [10]. Correlation functions in question have not been exploited in previous investigations of magnetic materials. They are spherical atomic multipoles with an even rank (quadrupoles in Fig. 1), with axial and magnetic discrete symmetries. Quantum mechanical selection rules in atomic physics forbid even rank multipoles in a J-manifold used by the pseudo-spin model. Magnetic space-groups we derive allow axial multipoles in the intensity of the specific Bragg spots chosen for investigation, and the exact angular anisotropy in a pattern is delineated. Interestingly, Dirac (polar) multipoles are forbidden in diffraction although they are allowed by Ir site symmetry.

Use of magnetic multipoles to encapsulate electronic degrees of freedom affords a means by which to move the knowledge of iridates forward by other techniques. Already, multipoles can be estimated with a program that is available for the interpretation of x-ray absorption and scattering experiments [24], while a different computational method has been exploited to estimate an exotic ordering of odd-rank multipoles in $URu_2Si_2$ [25].

**Acknowledgements** We thank P. Bourges for a pre-print of reference [10], additional data sets and valuable comments. G. van der Laan commented on early versions of the Letter. S. W. L. is grateful to P. Bargueño, L. C. Chapon, E. W. L. Grindrod, Sir Peter Knight and Svetlana Kozlova for assistance and guidance.
-----------------------------------------------------------------------------------------------------------

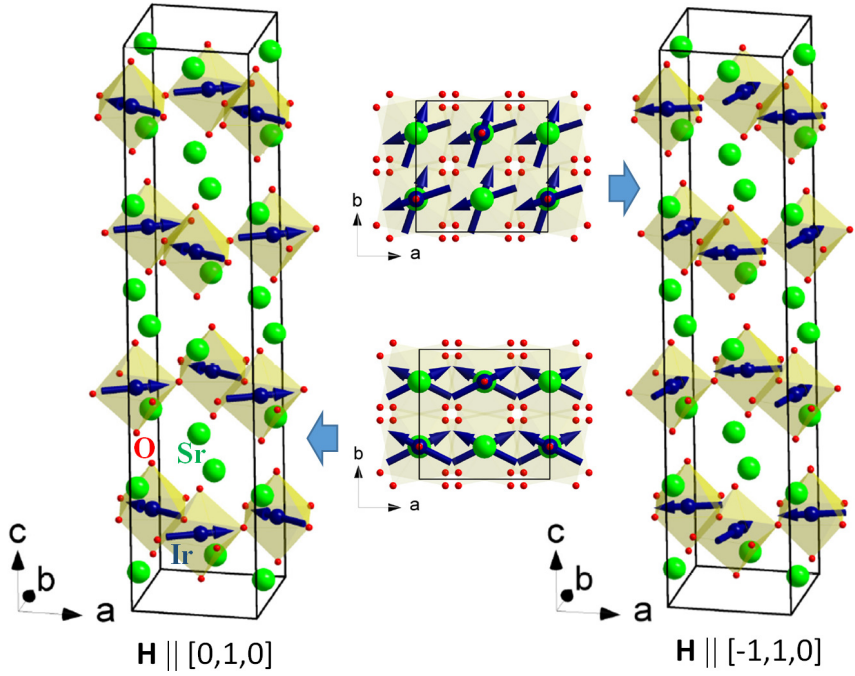

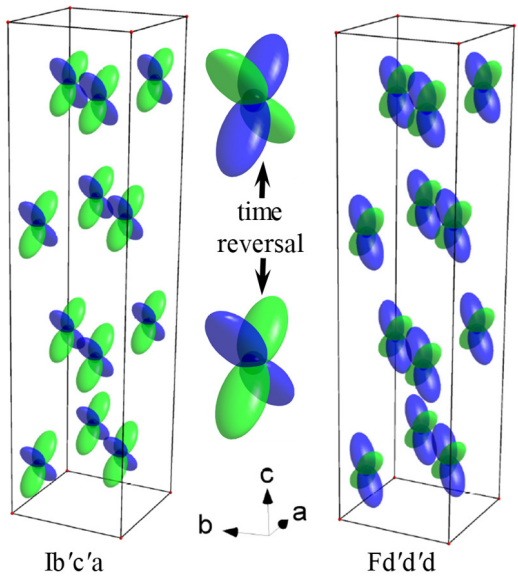

Fig. 1. Top panel; field-induced magnetic structures for the cases (I) & (II) when the magnetic field is applied along [0,1,0] (left) and [−1,1,0] (right), respectively. The corresponding magnetic space groups are (I) Ib′c′a and (II) Fd′d′d. Cell lengths $a = b \approx 5.484$ Å and $c \approx 25.804$ Å at 13 K [11]. Bottom panel; depiction of quadrupole moments $\langle T^2_{+2} \rangle''$.

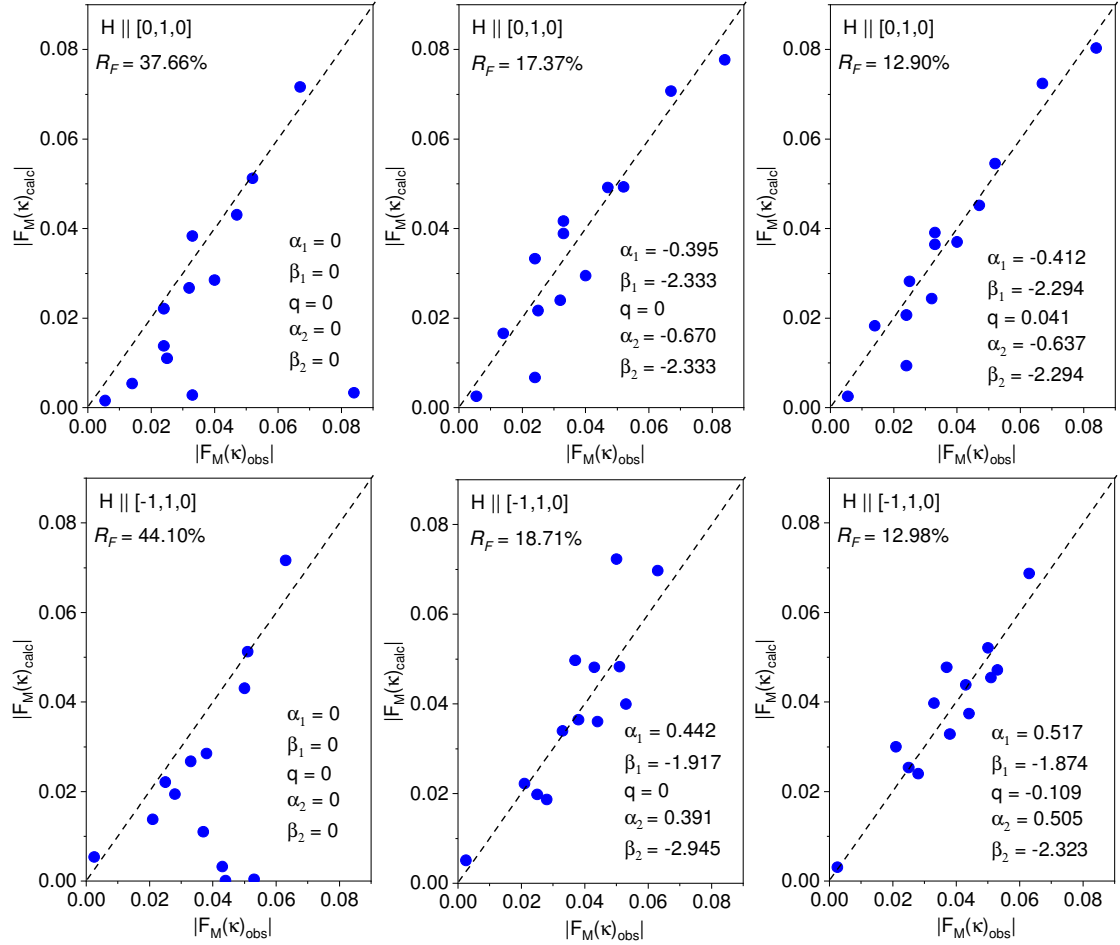

Fig. 2. Fits to data for the two field directions labelled (I) (top row) and (II) (bottom row) in the main text. Parameters determine multipoles $\langle T^2_{+2}\rangle'' = [q \langle j_2(\kappa)\rangle]$, $\langle T^3_0\rangle = \{\alpha_1 [\langle j_2(\kappa)\rangle + \beta_1 \langle j_4(\kappa)\rangle]\}$, and $\langle T^3_{+2}\rangle' = \{\alpha_2 [\langle j_2(\kappa)\rangle + \beta_2 \langle j_4(\kappa)\rangle]\}$. From left to right, fit to $F_M \approx [\langle (2\mathbf{S}+\mathbf{L})_\zeta\rangle \langle j_0(\kappa)\rangle]$ with $\langle (2\mathbf{S}+\mathbf{L})_\zeta\rangle = 0.08$, fit to (1) with q = 0 and, finally, fit with all three multipoles. Bragg diffraction data reported by Jeong *et al*. [10].

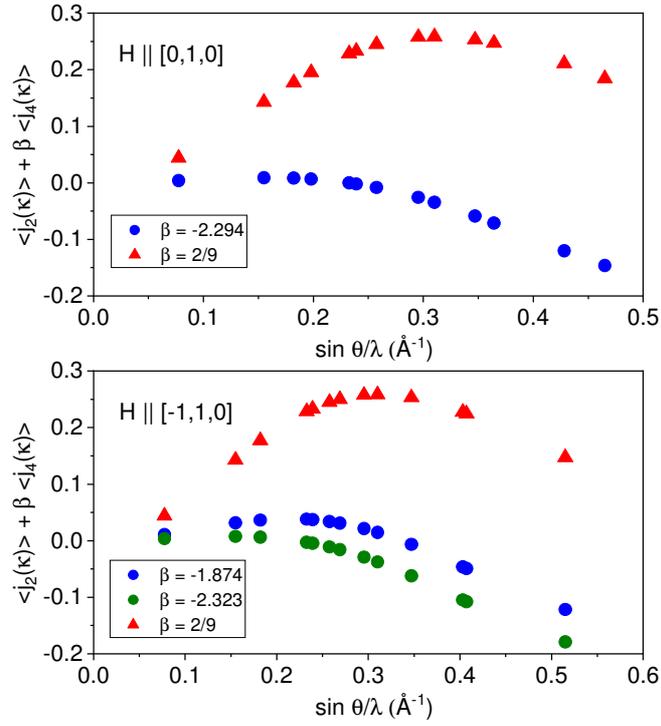

Fig. 3. $[\langle j_2(\kappa)\rangle + \beta \langle j_4(\kappa)\rangle]$ for various $\beta$ as a function of $\kappa/(4\pi) = \sin(\theta)/\lambda$ (Å$^{-1}$) determined by measured Bragg spots, as in Fig. 2. Red triangles $\beta = (2/9)$ appropriate for the J = 5/2 manifold used by the pseudo-spin model [8, 12]. Top panel field direction labelled (I); blue spots $\beta = -2.294$ inferred from data. Bottom panel case (II); blue (green) spots $\beta = -1.874$ ($-2.323$). Radial integrals $\langle j_n(\kappa)\rangle$ for Ir$^{4+}$ taken from reference [20] and $\langle j_2(0)\rangle = \langle j_4(0)\rangle = 0$.